# High Energy Solar Particle Events and Their Relationship to Associated Flare, CME and GLE Parameters

C. O. G. Waterfall[1] , S. Dalla[1] , O. Raukunen[2,3] , D. Heynderickx[4] , P. Jiggens[5] , and R. Vainio[3]

[1]Jeremiah Horrocks Institute, University of Central Lancashire, Preston, UK, [2]Aboa Space Research Oy, Turku, Finland, [3]Department of Physics and Astronomy, University of Turku, Turku, Finland, [4]DH Consultancy, Leuven, Belgium, [5]European Space Research and Technology Centre (ESTEC), Space Environment and Effects Section, Noordwijk, The Netherlands

**Abstract** Large solar eruptive events, including solar flares and coronal mass ejections (CMEs), can lead to solar energetic particle (SEP) events. During these events, protons are accelerated up to several GeV and pose numerous space weather risks. These risks include, but are not limited to, radiation hazards to astronauts and disruption to satellites and electronics. The highest energy SEPs are capable of reaching Earth on timescales of minutes and can be detected in ground level enhancements (GLEs). Understanding and analyzing these events is critical to future forecasting models. However, the availability of high energy SEP data sets is limited, especially that which covers multiple solar cycles. The majority of analysis of SEP events considers data at energies <100 MeV. In this work, we use a newly calibrated data set using data from Geostationary Operational Environmental Satellite-high energy proton and alpha detector between 1984 and 2017. Analysis of the SEP events in this time period over three high energy channels is performed, and SEP properties are compared to flare and CME parameters. In addition, neutron monitor (NM) observations are examined for the relevant GLE events. We find that correlations between SEP peak intensity and the CME speed are much weaker than for lower SEP energies. Correlations with flare intensity are broadly similar or weaker. Strong correlations are seen between >300 MeV data and GLE properties from NM data. The results of our work can be utilized in future forecasting models for both high energy SEP and GLE events.

**Plain Language Summary** Large eruptive events on the Sun can produce particles over a range of energies. The highest energy particles are the most hazardous and can quickly spread through space and reach Earth within minutes. The particles can interfere with satellite electronics and cause increased radiation doses to astronauts as well as pilots and flight passengers. Many studies exist of low energy particles, due to the increased number of spacecraft able to observe at these energies. However, the rarer high energy particles pose the most risks and are important to understand more. This study utilizes a new data set from a spacecraft that covers these most hazardous energies. This high energy particle data set is compared to features from the solar eruption, for example, the size of the associated solar flare. Strong relations are found between the high energy particles observed in space and their subsequent detection at ground level. This study can be used to improve forecasts of these events and improve our understanding and mitigation of these hazardous events.

## 1. Introduction

Recent decades have seen steady advances in science and technology that have led to a global reliance on satellites and associated technologies as well as an increase in space exploration. With these advances have come a need for a greater scientific understanding of relevant hazards and risks to these fields. One of these hazards that has the potential to disrupt and damage many areas of modern life is the occurrence of solar energetic particle (SEP) events. SEPs are associated with large solar eruptive events, such as flares and coronal mass ejections (CMEs). During these energetic and sudden phenomena, particles can be accelerated to several GeV. These SEPs travel through interplanetary space and can reach Earth on timescales as short as minutes following an eruption. Potential ramifications of these events include the disruption and even destruction of satellite electronics as well as increased radiation risk to air crew and astronauts. SEPs can also affect High Frequency radio communications in polar regions (Fiori et al., 2022). Indeed, space weather is not limited to Earth and needs to be considered for the habitability of any planet as well as any spacecraft launched outside Earth's protective magnetosphere. The ability to forecast and monitor SEP events, in particular the most hazardous high energy particles, is paramount to the advance of space weather research.

Many studies have analyzed how properties of SEP events relate to features of solar events (e.g., peak X-ray flare intensity, CME speed) to aid the refinement of space weather forecasting models. The first of these studies date back to the 1970s (Kahler et al., 1978; Van Hollebeke et al., 1975). Since then, numerous analyses have been performed as more data has become available. Gopalswamy et al. (2012) studied the ground level enhancement (GLE) events of solar cycle 23 and their associated solar eruption properties, and later the SEP events of the relatively quiet solar cycle 24 (Gopalswamy et al., 2014; Xie et al., 2019). Both of these solar cycle 24 studies focused on the role of CMEs in the generation of high energy SEP events. The solar cycle 23 study by (Gopalswamy et al., 2012) took a detailed look at CME parameters (e.g., height, speed, size) associated with the SEP events as well as their observed radio and X-ray features. They suggest that for GLE events, associated CME's from poorly connected regions have reached heights 66% larger than those from well-connected (W20-W90) longitudes at particle release time. Other analyses include an in-depth study by Dierckxsens et al. (2015). They examined events from solar cycle 23 and found numerous correlations between SEP, flare and CME properties. The correlations were derived for proton energies >10 and >60 MeV. Papaioannou et al. (2016) more comprehensively studied SEP events observed by Geostationary Operational Environmental Satellite (GOES) that occurred between 1984 and 2013 and also found consistent relationships between properties of the different events. The Papaioannou et al. (2016) study was conducted for proton energies >10, >30, >60, and >100 MeV. Some long standing notions in space weather are seen in these results, for example, the tendency for more SEP events to originate at Western longitudes. Generally speaking, these studies have found that large SEP events (as measured by both the peak flux and fluence) are associated with intense flares and fast CMEs. Papaioannou et al. (2016) found the strongest correlation to be between the CME speed and SEP peak flux and fluence.

Analysis of correlations between SEP and flare/CME parameters are used to refine and inform space weather forecasting models. For example, Marsh et al. (2015) used the correlations between peak flare X-ray intensity and SEP peak flux at >10 and >60 MeV from Dierckxsens et al. (2015) to scale the physical flux values predicted by their forecasting model, Solar Particle Radiation SWx (SPARX). The FORecasting Solar Particle Events and Flares tool uses statistical analysis of previous SEP events to calculate characteristics of an SEP event, for example, time of maximum peak flux, event duration, etc. (Anastasiadis et al., 2017). University of MAlaga Solar particle Event Predictor also uses correlations between flare and particle fluxes to predict onset of SEP events following flares of peak SXR intensity >C1.0 (Núñez, 2022).

While the field of space weather forecasting has grown significantly, many of these models and studies are focused on the lower energy end of the SEP proton energy range, typically on protons from ≈1 MeV to a few 10s of MeV. In the traditional NOAA definition, an SEP event occurs when a threshold flux of 10 proton flux units (p.f.u) for energies ≥10 MeV is exceeded, where 1 p.f.u = 1 particle cm$^{-2}$ s$^{-1}$ sr$^{-1}$. However, the highest space weather risk comes from the events capable of accelerating protons to >100 MeV. NOAA also issues alerts for ≥100 MeV protons that meet a 1 p.f.u threshold. Whilst events detected at >100 MeV are rarer, their occurrence poses the most risk. The sparse availability of high energy (>100 MeV) datasets limits the analysis of high energy event characteristics and thus our understanding of relativistic events. The analysis by Papaioannou et al. (2016) made use of differential proton fluxes from GOES observations at >100 MeV, with 123 of their 314 SEP events detected in this range. However, comprehensive analysis over multiple solar cycles at energies extending toward GeV energies are lacking. Bruno et al. (2018) studied a set of extreme SEP events observed by the Payload for Antimatter Matter Exploration and Light-nuclei Astrophysics (PAMELA) covering SEP intensities up to a few GeV but only observed between 2006 and 2016. Kühl et al. (2017) examined a set of 42 SEP events observed by the Solar and Heliospheric Observatory-Electron Proton Helium Instrument for >500 MeV between 1995 and 2005 but did not focus on comparisons with flare or CME parameters like previous studies. Instead, they examined neutron monitor (NM) count rates. Neutron monitor data are a valuable tool in studying these highest energy SEP events as many of them go on to cause GLE events on Earth (Oh et al., 2010). Oh et al. (2010) examined GOES data from 85 SEP events since 1986. They found good correlations between NM maximum increases and peak intensities (and fluences) across the GOES energy channels, with a pronounced threshold below which there is no GLE associated SEP event. For example, for the 350–420 MeV GOES channel the threshold peak flux is $4.16 \times 10^{-3}$ cm$^{-2}$ s$^{-1}$ sr$^{-1}$ MeV$^{-1}$ where 43 events were below this and none were GLEs. The full list of threshold values for each energy channel can be found in Table 7 of Oh et al. (2010). Investigating why some SEP events cause GLEs while others do not, is another important aspect of understanding these potentially hazardous events. One factor is the availability of particles accelerated to energies >500 MeV; required to generate the secondary particles that are detected by neutron monitors during a GLE.

One detector capable of high energy observations is the high energy proton and alpha detector (HEPAD) instrument onboard GOES (Onsager et al., 1996). GOES remains one of the longest operational spacecraft series and has been used in countless studies of large solar and space weather events (e.g., Lario et al. (2008)). Lario et al. (2008) examined GOES data from major SEP events during solar cycles 22 and 23 to explore the argument that a "streaming limit," that is, maximum peak intensity, exists for the prompt component of SEP events. Exceptions to this limit were found in both the P5 (39–82) and P7 (110–500 MeV) GOES channels. A streaming limit was first suggested by Reames and Ng (1998) to occur during the early phase of large SEP events, due to wave-particle interactions hindering the streaming of particles from the Sun. They determined this upper limit on the peak intensity for the higher energy channels of GOES/EPS (energetic particle sensor) observations as $1.2 \times 10^1$ and $6.5 \times 10^{-1}$ protons cm$^{-2}$ s$^{-1}$ sr$^{-1}$ MeV$^{-1}$ for P5 and P7, respectively. The limit is suggested to apply to particles in the prompt component, not those accelerated nearer to the observer, for example, in energetic storm particle (ESP) events (ESPs) where particle intensities can increase due to the passage of interplanetary shocks driven by CMEs.

Recently as part of the ESA HIERRAS project (Heynderickx et al., 2022), extensive work has been performed on GOES data to produce a new, calibrated and background subtracted data set that covers the time period 1984–2017, including data from the HEPAD instrument. This paper utilizes the new data set of re-calibrated GOES data covering 4 solar cycles to investigate correlations between properties of >350 MeV SEP events and those of the associated flares and CMEs. It uses three differential channels in the energy range between 300 and 750 MeV that have been interpolated from the re-calibrated GOES medium and high energy instrument channels. Comparisons with NM data for GLE events are also made to visualize differences and similarities between the two types of observations.

A description of the data set and the event selection is outlined in Sections 2 and 3, respectively. Comparison of the high energy data with flare and CME parameters are split into Sections 4 and 5. As GLEs are an important consequence of large space weather events, analysis of NM data for the GLEs in the SEP event list is also carried out in Section 6. The role of the heliospheric magnetic field is explored in Section 7. Finally, conclusions from the analysis of the SEP data and comparisons with other available data relating to the SEP events are discussed in Section 8.

## 2. Data Set

As described in Section 1, there is an unfortunate lack of datasets that describe relativistic particles in the near-Earth environment. In the preliminary stages of this study we examined what datasets were available and found many to have large data gaps, or limited to small and recent time periods. For example, PAMELA extends to high proton energies however it only covers events between 2006 and 2016. Interplanetary Monitoring Platform-8/Goddard Medium Energy (IMP-8/GME) data is available from the 1970s however has large data gaps during many large SEP events of interest. GOES is used in this study as it covers the largest time period, with the least data gaps, whilst still covering proton energies >300 MeV.

The data set used in this study was constructed within the ESA HIERRAS project. It is based on the SEPEM reference data set (version 2, RDS v2) (P. Jiggens et al., 2018; P. T. A. Jiggens et al., 2012), extended in energy with re-calibrated HEPAD data (Raukunen et al., 2020). The RDS v2 is based on proton data observed by the Synchronous Meteorological Satellite and GOES satellites which has been cleaned and cross-calibrated using IMP-8/GME data (Rodriguez et al., 2017; Sandberg et al., 2014). As the effective energies of the cross-calibrated channels vary between GOES satellites, the data have been re-binned into 11 logarithmically spaced energies between 5 and 289.2 MeV.

The HEPAD data have been re-calibrated using a "bow-tie" analysis of channel response functions (originally described by Van Allen et al. (1974)). The bow-tie analysis determines an optimal pair of effective geometric factor $G\Delta E$ and effective energy $E_{\text{eff}}$ for an instrument channel with a wide response in energy, so that the results are valid to a wide range of incident spectra. It is achieved by setting

$$R = \int_0^\infty j(E)G(E)\,dE = j(E_{\text{eff}})G\Delta E, \tag{1}$$

where $R$ is the observed counting rate, $j(E)$ is the true spectrum (assumed to be isotropic) and $G(E)$ is the calibrated channel geometric factor as a function of energy. Solving for $G\Delta E$ with a range of spectra (power laws with indices from −8.9 to −1.9 in case of HEPAD analysis) results in a family of curves for $G\Delta E$ as function of $E_{\text{eff}}$ resembling a bow-tie. The optimal pair of values can be found at the "knot" of the bow-tie. More details as well as results for all the HEPAD channels can be found in Raukunen et al. (2020).

**Table 1**
*Differential Proton Energy Channels of the Re-Calibrated Geostationary Operational Environmental Satellite Data Set ($cm^{-2}\ s^{-1}\ sr^{-1}\ MeV^{-1}$)*

| Channel | Central energy | Energy range |
|---|---|---|
| Ch12 | 347.8 | 289.2–418.3 |
| Ch13 | 503.0 | 418.3–604.9 |
| Ch14 | 727.4 | 604.9–874.7 |

The RDS v2 channels were extended with three high energy channels between 289.2 and 874.7 MeV (channels 12–14; see Table 1 for details). GOES channels P7, P9, and P10 were used in the re-binning for the added channels; the fluxes were calculated by log-log-interpolation of power laws between adjacent channels. To properly treat the high background of the HEPAD channels, background subtraction was performed on all channels before the energy re-binning. However, because of the logarithmic interpolation, all re-binned fluxes were set to zero whenever either of the adjacent background-subtracted GOES channel fluxes were zero, which may cause underestimation of fluxes and fluences in case of small events. The energies of the original GOES P7, P9, and P10 channels vary between GOES spacecraft. For P7 the energy channels are 110–500 (GOES 5–7), 165–500 (GOES 8–12), and 110–900 MeV (GOES 13–15). For P9 and P10 the energy channels are 435–555 and 555–760 MeV (GOES 6), respectively, and 420–510 and 510–700 MeV (GOES 8–13).

## 3. Event Selection

As described in the previous section, our data set spans the time period 1984–2017. This covers SEP events that occurred in solar cycles 22, 23, and 24 and part of 21. Papaioannou et al. (2016) carried out an analysis of SEP events between 1984 and 2013 up to 100 MeV. A comprehensive analysis of >300 MeV events and flare and CME characteristics over a similar time span has not been done before. Previously we have been limited by a small number of high energy events per solar cycle due to their rarity, and also a lack of complete high energy data sets.

We initially looked at existing SEP catalogs to compile a list of SEP events to check against the high energy data, for example, Papaioannou et al. (2016), Paassilta et al. (2017), https://umbra.nascom.nasa.gov/SEP/. The latter of these details the NOAA SEP list. The definition of NOAA events can lead to multiple events being listed as singular events. If the flux rises above and then falls below the ≥10 MeV 10 p.f.u threshold, an event is defined. If multiple flux increases occur over a short period of time, with no drop below the threshold, only one event is registered. Numerous SEP catalogs exist, however some, for example, SOHO/ERNE, cover much shorter time ranges than is necessary here. While some of the events were clearly defined in the high energy data, some events had very poorly defined (e.g., short and flat profiles with no clear rise or decay) or non-existent profiles. Additionally, some occurred at times of high quiescent background flux (i.e., around solar minimum in 1998) so their profiles are not defined. To classify as an event in our study the flux had to be non-zero for a >3 hr continuous period. The event end is defined as when the period of non-zero flux ends.

This selection produced 49 separate SEP events in the high energy data. We further reduced this to 42 events by excluding those which had a flare behind the limb. This was done as there is insufficient information on the flare location and magnitude from GOES observations, which are important in our later comparison of flare and SEP event properties in Section 4. The full list of 42 SEP (29 GLE and 13 non-GLE events) included in our study is listed in Table 2. The non-GLE events includes 2 sub-GLEs: 27 January 2012 and 7 March 2012. A sub-GLE is defined as an SEP event detected by high-altitude polar neutron monitors but with no significant count rate increase at sea-level stations (Poluianov et al., 2017). In total there are 1, 17, 19, and 5 events from solar cycles 21–24 respectively. The list of behind limb flares that meet the event criteria are listed in Table 3.

Some SEP events are not included in this study, either due to having a non-zero flux for a short time (<2 hr) or from being removed entirely due to the background subtraction. These include 30 November 1989, 23 November 2001, 23 January 2012, and are all seen in the original, raw, HEPAD data. An example of the lower fluxes induced in the background subtraction is illustrated in the series of GLEs during October 1989 in Figure 1.

Parameters of the events in Table 2 which were calculated are:

- Peak flux
- Event onset time
- Event duration

**Table 2**
*List of Solar Events That Produced High Energy Solar Energetic Particle Enhancements Meeting the Criteria Described in Section 3*

| GLE no. | Flare date | Flare class | Flare location | CME speed (km s$^{-1}$) |
|---|---|---|---|---|
|  | 14 March 1984 | M2.0 | S12W42 |  |
| 40 | 25 July 1989 | X2.6 | N25W84 |  |
|  | 12 August 1989 | X2.6 | S16W37 |  |
| 41 | 16 August 1989 | X20 | S18W84 |  |
| 42 | 29 September 1989 | X9.0 | S26W90 | 1,828 |
| 43 | 19 October 1989 | X13.0 | S27E10 |  |
| 44 | 22 October 1989 | X2.9 | S27W31 |  |
| 45 | 24 October 1989 | X5.7 | S30W57 |  |
| 46 | 15 November 1989 | X3.2 | N11W26 |  |
| 47 | 21 May 1990 | X5.0 | N35W36 |  |
| 48 | 24 May 1990 | X9.3 | N33W78 |  |
|  | 23 March 1991 | X9.4 | S26E28 |  |
|  | 13 May 1991 | M8.2 | S09W90 |  |
| 51 | 11 June 1991 | X12 | N31W17 |  |
| 52 | 15 June 1991 | X12 | N33W69 |  |
|  | 30 October 1991 | X2.5 | S08W25 |  |
| 53 | 25 June 1992 | X3.9 | N09W67 |  |
|  | 30 October 1992 | X1.7 | S22W61 |  |
| 55 | 06 November 1997 | X9.4 | S18W63 | 1,556 |
| 56 | 02 May 1998 | X1.1 | S15W15 | 938 |
| 57 | 06 May 1998 | X2.7 | S11W65 | 792 |
| 58 | 24 August 1998 | X1.0 | S35E09 |  |
| 59 | 14 July 2000 | X5.7 | N22W07 | 1,674 |
|  | 08 November 2000 | M7.4 | N10W77 | 1,738 |
| 60 | 15 April 2001 | X14.4 | S20W85 | 1,199 |
|  | 24 September 2001 | X2.6 | S16E23 | 2,402 |
| 62 | 04 November 2001 | X1.0 | N06W18 | 1,810 |
| 63 | 26 December 2001 | M7.1 | N08W54 | 1,446 |
|  | 21 April 2002 | X1.5 | S14W84 | 2,393 |
| 64 | 24 August 2002 | X3.1 | S02W81 | 1,913 |
| 65 | 28 October 2003 | X17 | S16E08 | 2,459 |
| 66 | 29 October 2003 | X10 | S15W02 | 2,029 |
| 67 | 02 November 2003 | X8.3 | S14W56 | 2,598 |
| 68 | 17 January 2005 | X3.8 | N15W25 | 2,094 |
| 69 | 20 January 2005 | X7.1 | N14W61 | 3,256 |
|  | 06 December 2006 | X6.5 | S05E64 |  |
| 70 | 13 December 2006 | X3.4 | S05W23 | 1,774 |
| sub | 27 January 2012 | X1.7 | N27W71 | 2,508 |
| sub | 07 March 2012 | X5.4 | N17E27 | 2,864 |
| 71 | 17 May 2012 | M5.1 | N11W76 | 1,582 |

- Event fluence (calculated as the time integrated flux between event onset to end)
- Event time to maximum (defined as the time between onset and peak flux)

These parameters are all used in the analysis of the high energy events detailed in this work. First, these properties are studied alone, then in later sections the SEP event properties are compared to those of the associated flares, CMEs and NM data.

### 3.1. Properties of High Energy SEP Events

It is interesting to compare different features of the GLE and non-GLE populations. Of the 42 events in our list 29 are GLE events and 13 are non-GLE, showing that an event at these energies is likely associated with a GLE but not necessarily so. 20/29 of the GLE events have times to maximum (calculated as time between first non-zero flux and peak) of less than 2 hr (and 3/13 non-GLE events). Median time to max for GLE events is 85 min, 410 min for non-GLE events. Plotting the peak fluxes versus the fluences of the events in Figure 2 reveals strong correlations between the logarithmic values, with Pearson coefficients of 0.89, 0.85, and 0.89, respectively. However, for a given peak flux, the associated fluence can vary by an order of magnitude. This relationship remains strong even for Ch14 where there are fewer events. Of the events in Ch14, only 2 non-GLEs remain: 12 August 1989 and 23 March 1991. A similarly close relationship between the peak intensities and event fluences has previously been seen for SEP events (Kahler & Ling, 2018). However, the Kahler and Ling (2018) study focused on lower energies (>10 MeV protons). Therefore, it is useful to see the relationship between these parameters at higher energies presented in Figure 2.

The peak flux is further examined in Figure 3. The histograms show the distribution of peak fluxes in the three energy bands for all events. Lario et al. (2008) investigated the existence of a possible upper limit on SEP peak fluxes, the so-called "streaming limit." While they found a few events with peak intensities above the streaming limit, most were below it. This was suggested to be due to either a lack of events and data to confirm the limit, or the streaming limit itself. The streaming limit from Lario et al. (2008) is given for GOES P5 (39–82) and P7 (110–500 MeV) channels, not any of the high energy channels studied here. The distribution of events in Figure 3 does not show evidence for existence of an upper limit, however the number of events is small and studies with more SEP events is needed. All of our peak fluxes are taken from the prompt component of the SEP event, not any ESP related peak (see Section 4.1.1). The largest flux seen in these results is for GLE 69, an extremely prompt GLE in GOES and NM data. This was one of the SEP events from (Lario et al., 2008) that exceeded the suggested streaming limit in the P7 channel. Lario et al. (2008) suggested that this event surpassed the given streaming limit either due to a "magnetic bottleneck" from a prior interplanetary CME or an insufficient wave intensity to restrict the streaming past the given limit. The smallest peak flux in Ch12 is for the SEP event on 22 May 2013. GLE 57 (6 May 1998) has the smallest fluence in our sample. In contrast, the event with the largest fluence was GLE 43 (19 October 1989). The distribution of the events in Figure 3 vary over the three high energy channels. High peak flux events are rarer, with fewer events seen as the peak flux increases in all three channels. There are more events with

**Table 2**
*Continued*

| GLE no. | Flare date | Flare class | Flare location | CME speed (km s$^{-1}$) |
|---|---|---|---|---|
| | 22 May 2013 | M5.0 | N15W70 | 1,466 |
| 72 | 10 September 2017 | X8.2 | S08W88 | 3,163 |

*Note.* "sub" indicates sub-GLE events.

lower peak fluxes for Ch12. However, these smaller events are lost at higher energies, leaving a peak toward higher fluxes in Ch14. In order to understand these events further, features of the associated solar event are now compared to these SEP properties.

## 4. Solar Flare Parameters

This section examines the parameters of the solar flares associated with the events in Table 2. Relevant parameters include the flare location, flare magnitude and temporal behaviors (time to max, duration, etc.). The flare parameters are primarily obtained from the NOAA GOES Solar Proton Event (SEP) List (https://umbra.nascom.nasa.gov/SEP/seps.html) and cross checked with other lists (e.g., Cane et al. (2010) and Papaioannou et al. (2016)). Comparisons are also made between properties of the SEP event at high energies to those of the associated solar flares.

Figure 4 shows histograms of the magnitude (left) and longitude (right) of the flares associated with GLE events (red) and other SEP events (blue) in our event list. There is a clear tendency for higher magnitude flares to be associated with GLE events. In general, the high energy SEP events in our list are all associated with strong flares. Only eight events have flares less than X1.0 (4 GLEs and 4 SEP). For the flare magnitude histogram, a *t*-test suggests that there is a statistically significant difference between the GLE and SEP populations ($p$-value < 0.05).

The second histogram in Figure 4 shows a tendency for GLE events (and high energy SEP events in general) to be associated with Western longitude flares, as well established in literature, due to the Sun-Earth magnetic connectivity. However, Eastern GLEs are not unheard of. GLE 43 (19 October 1989) and 65 (28 October 2003) had Eastern longitudes and significant increases in both GOES and NM data. No extreme Eastern events (E90–E20) have caused a GLE however, despite large flares and fast CMEs. Not all Western longitude events lead to GLE events, even when associated with large flares and fast CMEs. This suggests that while the East/West location of these events is an important indicator, the position on the Western hemisphere itself cannot alone be used as a factor in determining whether an event will lead to a GLE. Whilst only the flare longitude has been considered here, this is also true for CMEs, where even a well-connected fast CME may not lead to an SEP event (as seen for some solar cycle 23 events (Gopalswamy et al., 2012)). For the flare longitude histogram, a *t*-test suggests that there is no statistically significant difference between the GLE and SEP populations ($p$-value > 0.05). Further analysis of this flare data is done in the following subsection, by comparing it to high energy data and exploring other possible observable variables in these events.

### 4.1. Comparison With High Energy Data

Analysis of SEP events has more commonly been performed at energies <100 MeV. One such analyses by Dierckxsens et al. (2015) examined the relationship between the flare magnitude and peak fluxes over different energies. The correlation coefficients from this analysis, as well as for Dierckxsens et al. (2015) and Papaioannou et al. (2016) are listed in Table 4. An important caveat with these SEP studies is that they all present the correlations between logarithmic values of the parameters, for example, the log of both the peak SEP flux and peak SXR flare intensity. This often leads to an overestimate of the strength of the relationship between the parameters. For the smaller values of the coefficients any correlation can vanish entirely when the true values are compared. However, we report the correlation coefficients for logarithmic relationships here to maintain consistency with previous studies.

Dierckxsens et al. (2015) used integral proton fluxes while both the Papaioannou et al. (2016) and this analysis use differential proton fluxes. We have extended this analysis to our high energy channels with the results shown in Figure 5. The three subplots show the results for the Ch12, 13, and 14 peak flux comparison with peak X-ray flare intensity for all events in Table 2. There are fewer data points than in the Dierckxsens et al. (2015) analysis as higher energy events are rarer. The low energy analysis also has a wider range of flare magnitudes, as expected from the lack of high energy events with <M2 flares on the disk. The correlation coefficients for Ch12, 13, and 14 in Figure 5 are 0.55, 0.39, and 0.28, respectively, as listed in Table 4. These decline with increasing energy, unlike in Dierckxsens et al. (2015), and are generally smaller than coefficients from lower energy analysis. In contrast, the correlations seen in Papaioannou et al. (2016) do not increase or decrease significantly with energy, with both >10 and >100 MeV having correlation coefficients of 0.49. The errors for all new coefficients in Table 4 are all large, due to the small number of high energy events.

**Table 3**
*Solar Energetic Particle Events That Are Present in the High Energy Channels but Occurred Behind the Limb*

| GLE no. | Flare date | Flare class | CME speed | Flare location |
|---|---|---|---|---|
| 39 | 16 February 1984 | | | >W90 |
| 49 | 26 May 1990 | X1.4 | | N33W104 |
| 50 | 28 May 1990 | C9.7 | | N33W120 |
| 54 | 02 November 1992 | X9.0 | | S25W100 |
| 61 | 18 April 2001 | C2.2 | 2,465 | >SW90 |
| | 16 August 2001 | | 1,575 | >SW90 |
| | 06 January 2014 | C2.1 | 1,402 | >W90 |

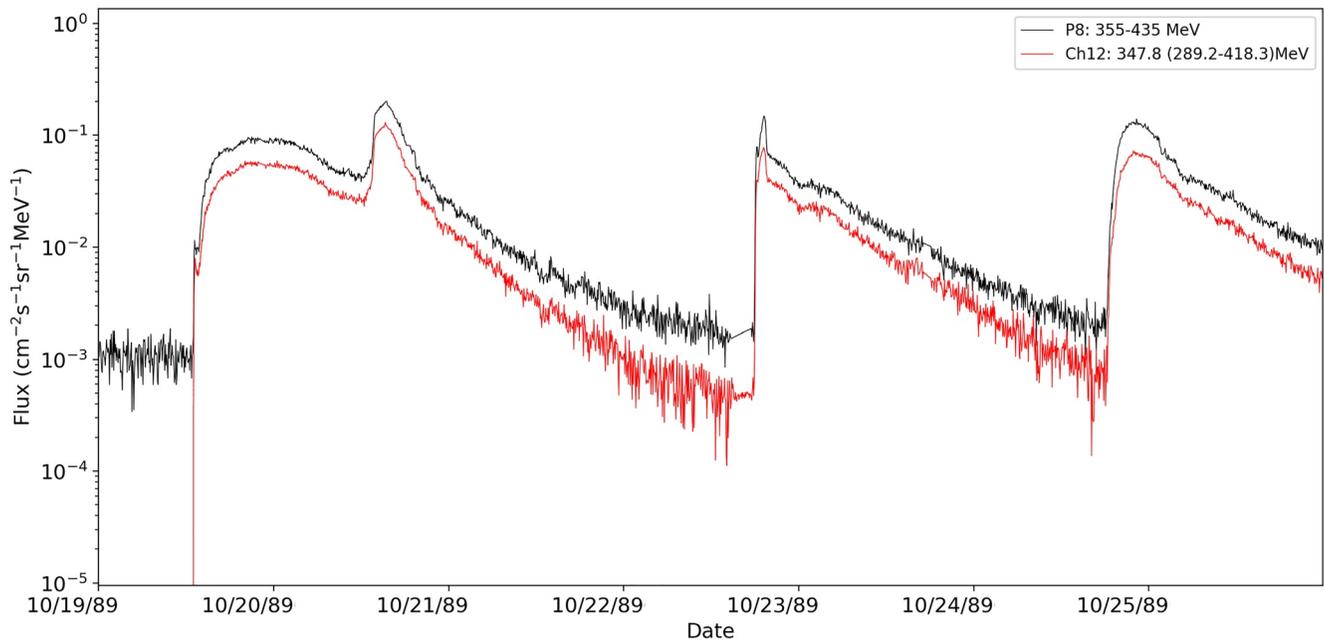

**Figure 1.** Original, uncalibrated high energy proton and alpha detector data (P8 channel: 355–435 MeV) shown in black and calibrated Ch12 data in red for the ground level enhancements 43, 44 and start of 45 during October 1989.

Fluences are plotted versus the flare magnitudes in upper panels of Figure 5. The correlations for the fluence are all stronger, with the largest coefficient for Ch12 at 0.62. These coefficient values are more comparable to those obtained from lower energy analysis by Papaioannou et al. (2016). Generally, the larger flares are associated with SEP events with larger fluences. This was also found by Papaioannou et al. (2016) in their lower energy study.

The analysis here suggests that there is a moderate relationship between the flare and event parameters even at high energies up to 500 MeV (when the correlation coefficient is calculated for the logarithmic values). However this correlation becomes much weaker for peak flux and fluence at energies higher than this. Previous studies, such as Dierckxsens et al. (2015), only considered solar cycle 23 events whereas our data set covers all of solar cycle 22–24. However, using only one solar cycle drastically decreases our already smaller data set. Indeed, in our sample we only have five events in solar cycle 24. Other parameters of the high energy SEP data were also compared with flare parameters, for example, event duration and decay time but no significant correlations were obtained.

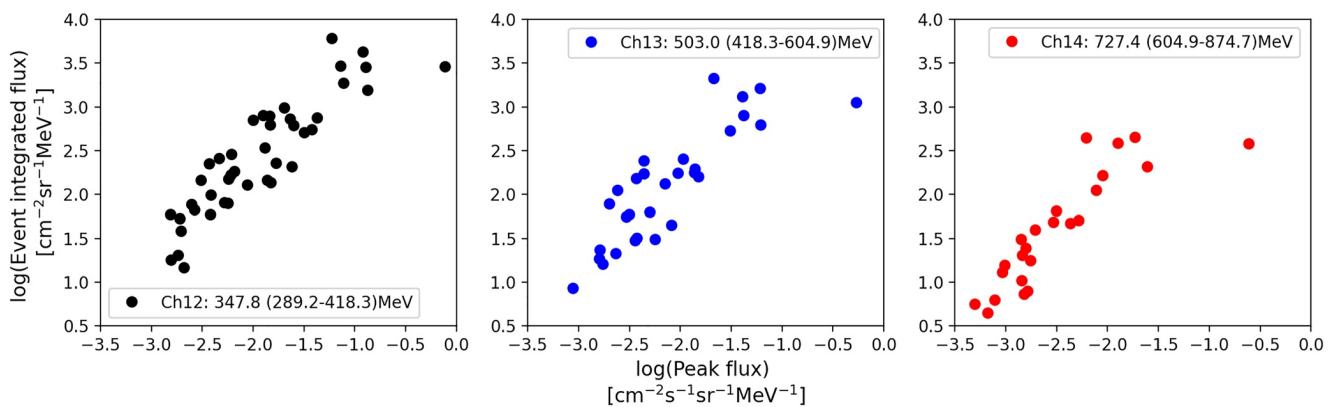

**Figure 2.** Plot of the logarithm of the fluence versus peak flux for all high energy solar energetic particle events observed. From left to right: Ch12, 13, and 14 with energies listed in Table 1.

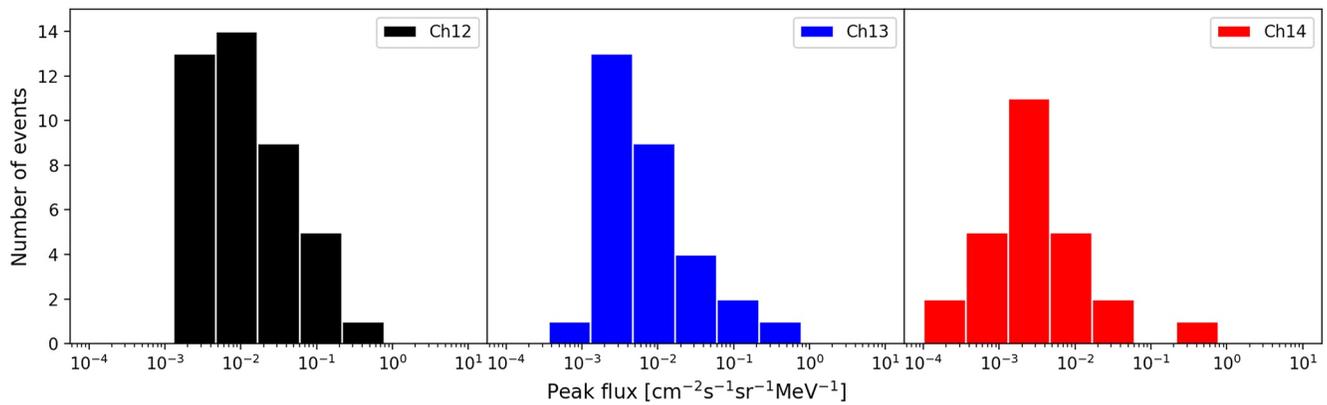

**Figure 3.** Histogram showing the distribution in peak flux values of all solar energetic particle events in the three energy channels (12, 13, and 14) from left to right.

#### 4.1.1. ESP Events

Some of the events in Table 2 are associated with ESP events in interplanetary space. Some of these ESP events appear in the data as secondary peaks. One notable example of a secondary peak is GLE 43, 19 October 1989, which had a large increase a day later that was higher than the initial peak. This is remarkably seen in all three high energy channels (e.g., Ch 12 in Figure 1). As mentioned in Section 3, this event had the largest fluence in our set, primarily because of this secondary rise. This feature is unique to GLE 43 and has been discussed in previous literature, for example, Lario and Decker (2002). There has been some speculation as to whether this ESP event was a result of local shock acceleration or from a complex plasma and magnetic field environment induced from prior activity (Cane & Richardson, 1995). We do not investigate the possible ESP nature of this peak further here, and treat it as a general secondary feature of the event, taking the initial maximum as our peak value. For the other events where an ESP peak does appear, it is always smaller than the initial value.

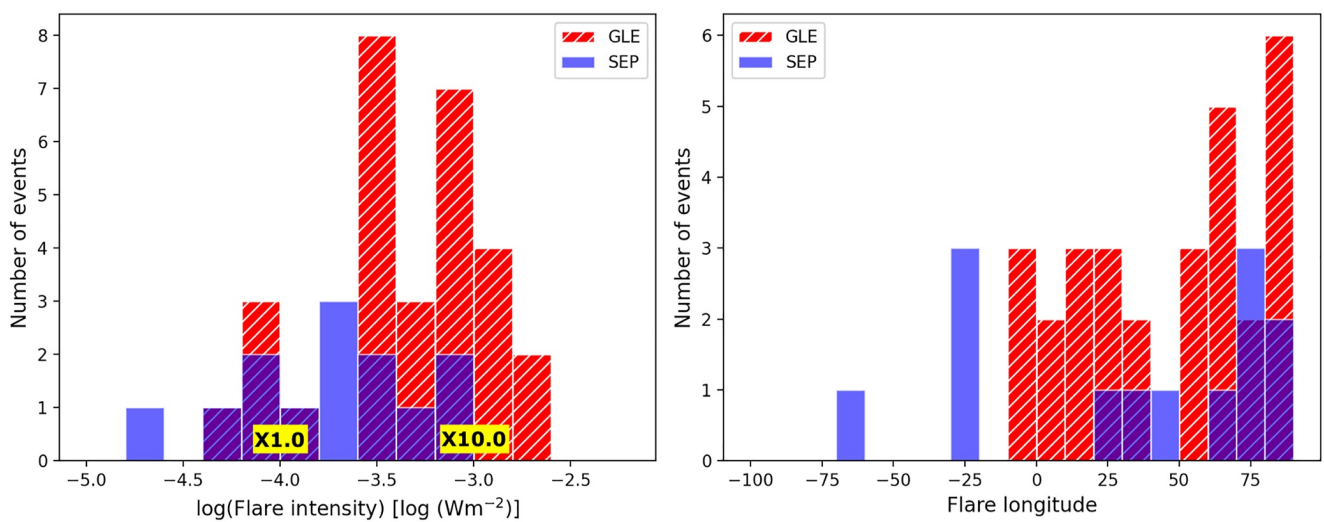

**Figure 4.** Histograms of flare magnitude (left) and longitude (right) of the events from Table 2. Red dashed and blue solid bars denote the ground level enhancement (GLE) and non-GLE events, respectively. Darker blue regions are solar energetic particle events overlapping the GLE events.

**Table 4**
*Correlation Coefficients From Previous Low Energy Solar Energetic Particle Analysis (Dierckxsens et al. (2015) and Papaioannou et al. (2016)) Compared With Results From Our Geostationary Operational Environmental Satellite Analysis*

| Energy | log$_{10}$(Peak SXR flare intensity) versus | | CME velocity versus | |
|---|---|---|---|---|
| | log$_{10}$(Peak SEP flux) | log$_{10}$(SEP fluence) | log$_{10}$(Peak SEP flux) | log$_{10}$(SEP fluence) |
| >10 MeV | 0.49[a], 0.55[b] | 0.48[a] | 0.57[a], 0.56[b] | 0.60[a] |
| >30 MeV | 0.47[a] | 0.42[a] | 0.51[a] | 0.53[a] |
| >60 MeV | 0.48[a], 0.63[b] | 0.41[a] | 0.44[a], 0.40[b] | 0.45[a] |
| >100 MeV | 0.49[a] | 0.42[a] | 0.40[a] | 0.46[a] |
| Ch12 | 0.55 ± 0.13 | 0.62 ± 0.12 | 0.26 ± 0.21 | 0.41 ± 0.20 |
| Ch13 | 0.39 ± 0.17 | 0.46 ± 0.17 | 0.14 ± 0.26 | 0.16 ± 0.25 |
| Ch14 | 0.28 ± 0.20 | 0.44 ± 0.18 | 0.28 ± 0.29 | 0.22 ± 0.29 |

*Note.* Top section shows coefficients between the log of the SEP peak flux (and fluence) and peak SXR flare intensity for energies studied by
[a]Papaioannou et al. (2016), [b]Dierckxsens et al. (2015) and Ch12, 13, and 14 used in this study. Lower section shows coefficients between the CME velocity and SEP peak flux (and fluence) for the same energies

## 5. CME Parameters

Similarly to the lower energy analysis performed by Dierckxsens et al. (2015), we have also examined the properties of the CMEs associated with SEP events. However, we are again limited by a lack of data. Many of our events originated in the active solar cycle 22, for which we have no (or very limited) CME data. For later solar cycles there is data available, but the number of high energy SEP events during these periods is small (<25 in 20 years). We use the CME plane of sky velocity in this study to compare with the high energy data, taken from Papaioannou et al. (2016) and the SOHO/LASCO CME catalog (Gopalswamy et al., 2009). The CME speed for GLE 42 was taken from Moraal and Caballero-Lopez (2014). Information on the width of the CMEs is available, but they are nearly all halo CMEs in this instance (the exceptions are: 06 May 1998, 08 November 2000, 15 April 2001, and 26 December 2001). It should be noted that reported speeds for halo CMEs often have a greater degree of uncertainty than non-halo CMEs, due to projection effects. This uncertainty is applicable to other analyses of CME and SEP relationships, including those by Dierckxsens et al. (2015) and Papaioannou et al. (2016), where the majority of SEP events are associated with halo CMEs. For example, 106/158 SEP events are associated with halo CMEs in the study by Papaioannou et al. (2016).

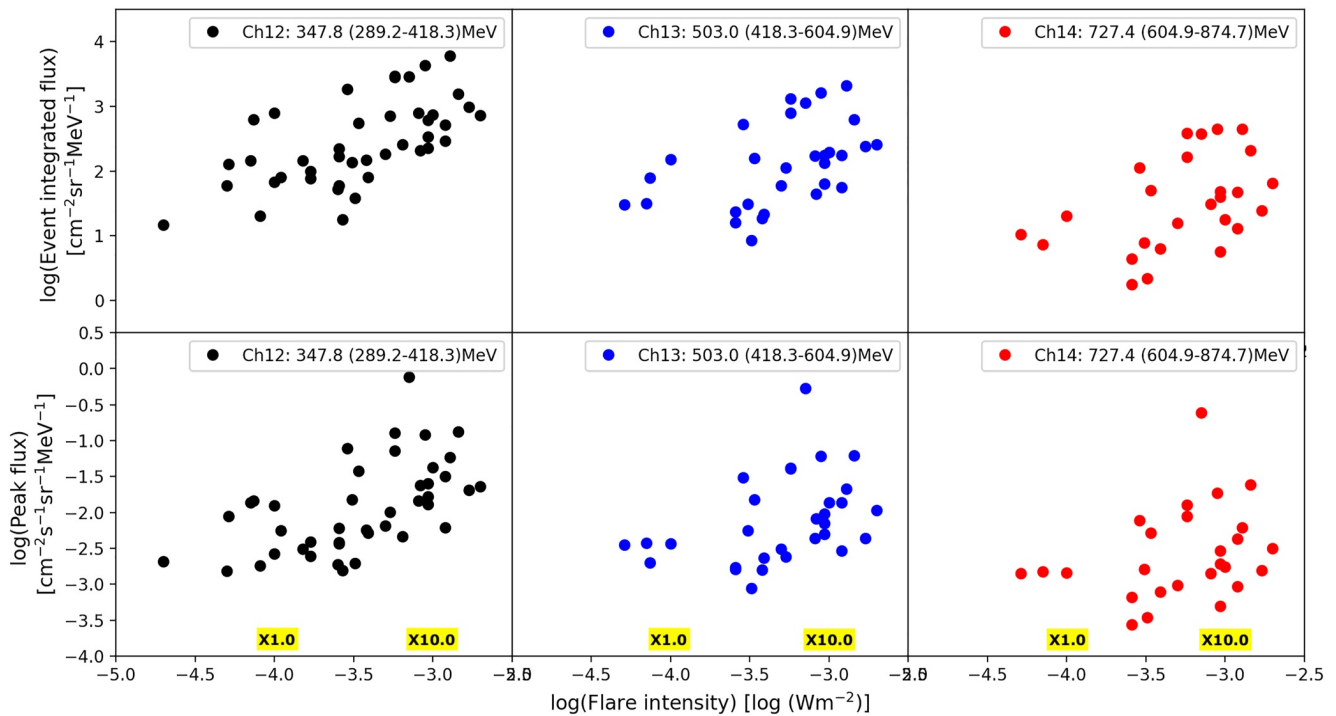

**Figure 5.** Plot of the logarithm of the proton peak flux (bottom panels) and logarithm of the fluence (top panels) versus logarithm of the peak X-ray flare intensity for the three high energy channels. The *y*-axes span equal orders of magnitude in each row.

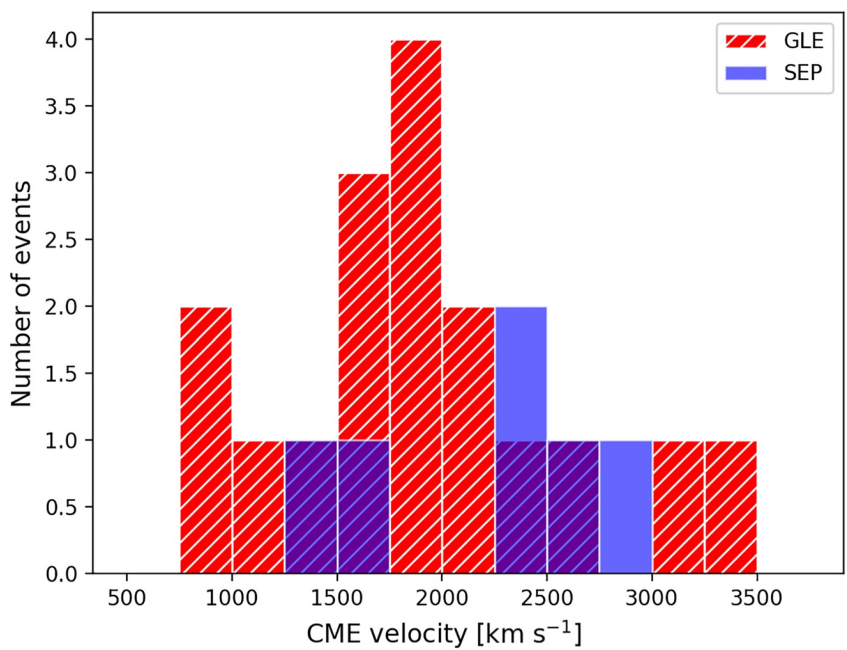

**Figure 6.** Histogram of coronal mass ejection velocities (km s$^{-1}$) for solar energetic particle (blue) and ground level enhancement (red) events from Table 2.

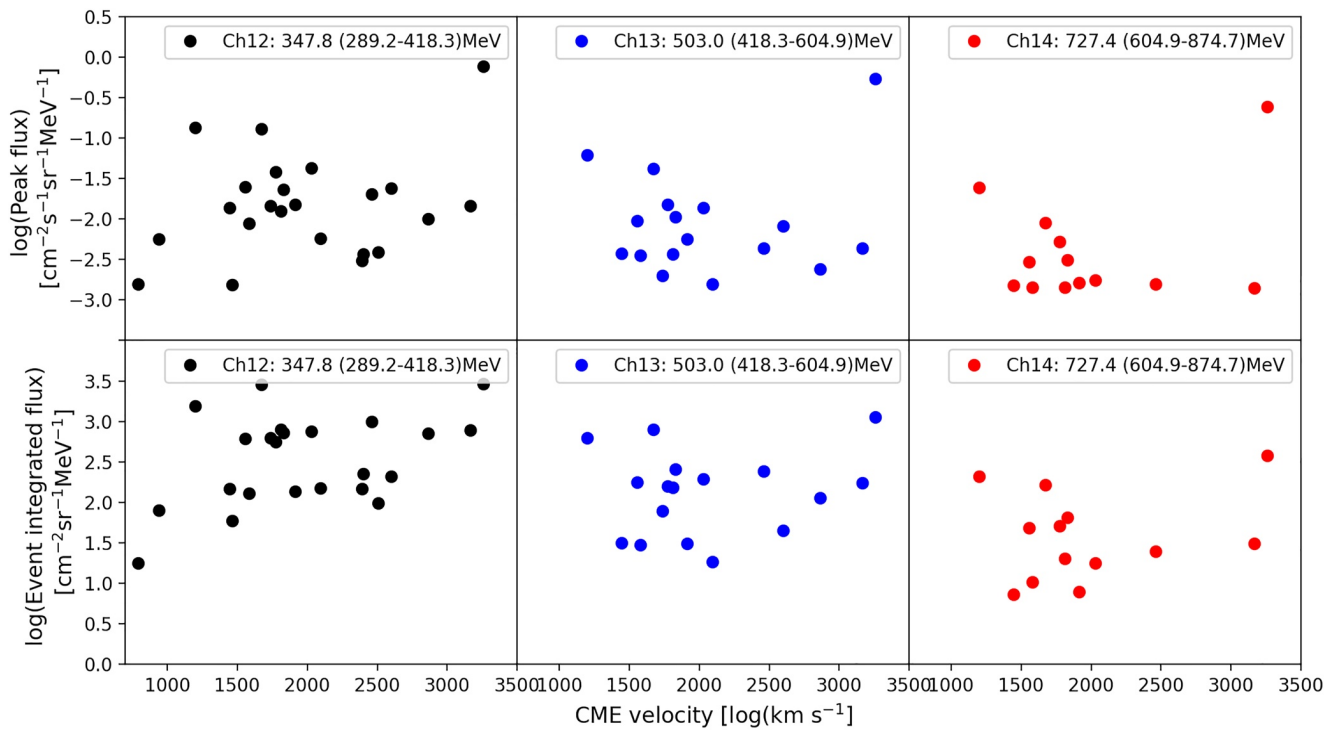

**Figure 7.** Plot of the logarithm of the peak proton flux (top panels) and logarithm of the event integrated flux (bottom panels) versus coronal mass ejection velocity (km s$^{-1}$) in the three high energy channels.

Figure 6 shows the distribution of CME velocities for our 17 GLE and 6 non-GLE events. All our events have fast CMEs, with the slowest speed reported for GLE 57 at 792 km s$^{-1}$. Fifty-three percentage of our GLEs (for which we have CME data available) have velocities between 1,500 and 2,500 km s$^{-1}$. There are too few data points for the non-GLE events to definitively comment on whether there are any differences between the populations. Indeed, a *t*-test performed on the CME velocity histogram suggests that there is no statistically significant difference between the GLE and SEP populations (*p*-value > 0.05).

We plot the CME velocity versus the peak flux and fluence from the three channels in Figure 7. The low energy analysis of Dierckxsens et al. (2015) reports correlation coefficients of 0.56 and 0.40 for >10 and >60 MeV, respectively. For Ch 12–14 we obtain coefficients of 0.26, 0.14, and 0.28, respectively. These values suggest a much weaker correlation. A faster CME does not produce a larger peak flux in any channel. This could be due to many reasons, including but not limited to: experimental error and differing methods used to calculate the CME velocity (this study uses plane of sky velocities), limited statistics due to lack of data from older events, transport effects playing a larger role at higher energies (e.g., gradient, curvature and heliospheric current sheet (HCS) drifts might influence higher energy particles more, distributing them more throughout the heliosphere). Correlations between CME speed and peak particle fluxes have been widely reported at lower energies, for example, Kahler (2001) and Bruno and Richardson (2021). Bruno and Richardson (2021) examined SEP events observed by STEREO and PAMELA at energies 10–130 MeV in the development of an empirical model for predicting SEP peak fluxes and fluences. Like Dierckxsens et al. (2015), they also found the correlation coefficient to decrease with increasing energy.

Similar results are seen in the bottom panels of Figure 7, plotting the speed of the CME versus the fluence at high energies. The Pearson correlation coefficients are again weak: 0.41, 0.16, and 0.22 for Ch12, 13, and 14, respectively. Comparing the fluence to both the flare and CME properties gives similar or stronger correlations than when the peak flux is used. This is also seen in Papaioannou et al. (2016), where the Pearson value is typically higher for a given energy when a comparison is made between an event parameter and the fluence, as opposed to peak flux. However, the correlations in Papaioannou et al. (2016) are more significant than those here. For example, they find a reasonable correlation (0.60 for >10 and 0.46 for >100 MeV) between fluence and CME velocity and suggest that faster CMEs (>1,000 km s$^{-1}$) are associated with larger fluences. Our corresponding correlations are weak at all energies and a similar conclusion cannot be drawn.

## 6. Neutron Monitor Parameters

As is clear in Table 2, many of the GOES high energy events are associated with GLE events. In this section we compare some of the properties of NM data to our data set to explore any relationships that exist between both types of observation. Figure 8 shows the Ch12 and Ch14 peak flux (top panels) and fluence (bottom panels) versus the de-trended maximum NM percentage increase and NM fluence. To obtain the maximum increase, data from the network of NMs were examined, obtained from https://gle.oulu.fi. De-trended refers to data from this international GLE database that have been corrected to account for the fluctuations from the galactic cosmic ray background (Usoskin et al., 2020). While the majority of the peak increases are found from the South Pole neutron monitors (filled circles in Figure 8), if a maximum occurs at a different NM that value is used. The fact that the largest increases per event tend to be at the South Pole neutron monitors is expected as these monitors are situated at high altitudes and have a low cut-off rigidity. Smaller increases are more common at stations closer to sea-level at non-polar latitudes that have to compete with geomagnetic shielding as well as atmospheric shielding. Varying cut-off rigidities can influence the detection and intensity of GLEs together with other factors such as Forbush decreases (e.g., GLE 72 occurred toward the end of a Forbush decrease where count rates were still suppressed (Hubert et al., 2019), and GLE 67 occurred during a Forbush decrease following GLE 66 (Mishev et al., 2021).

The peak flux and peak NM increase in Figure 8 are very strongly correlated for all high energy channels. For Ch12 and Ch14 the correlation coefficients are 0.879 and 0.925, respectively. In similar analysis of SEP events in GOES data, Oh et al. (2010) found corresponding coefficients of 0.874 and 0.867 for P8 and P10, respectively. Both sets of analysis independently show a strong correlation between the higher energy channels and the maximum increase from neutron monitors. This correlation improves slightly from Ch12 to Ch14 in this analysis, and decreases slightly between P8 and P10 (the most similar bands to ours) in the analysis by Oh et al. (2010). This relation holds even for the most extreme GLE, GLE 69. This GLE can be seen in the top right of both plots, with a significantly larger peak flux and NM increase than the others. A Ch13 panel is excluded from Figure 8 as the results were very similar to Ch12.

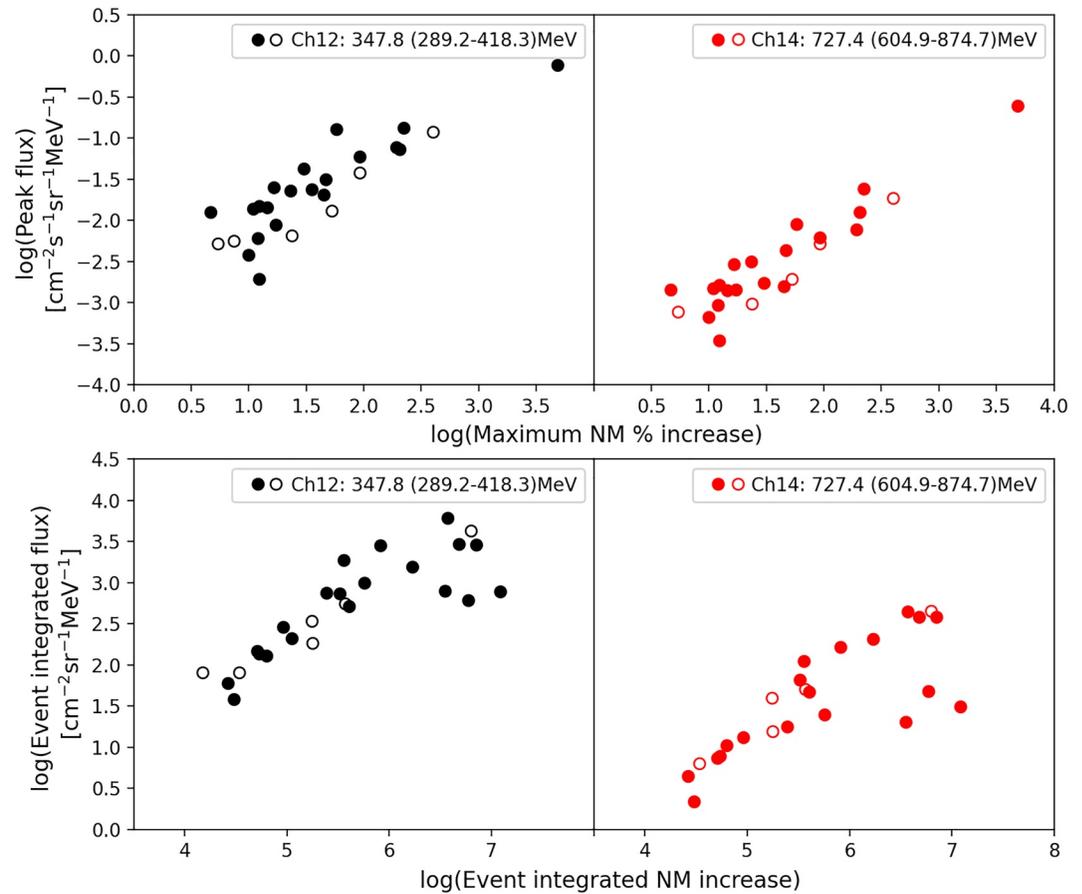

**Figure 8.** Plot of the Geostationary Operational Environmental Satellite (GOES) peak flux versus maximum neutron monitor (NM) percentage increase for ground level enhancements in our event list (top panels) and the GOES event fluences versus event integrated intensities of the NM events (bottom panels) in Channels 12 and 14. Unfilled circles indicate events where the NM data was not taken from South Pole monitors.

Similarly, a good correlation exists in Figure 8 (bottom panels) between the fluences of the GOES and NM events. For Ch12 and 14 the coefficient is 0.853 and 0.793, respectively. There are fewer GLEs included in the lower panels of Figure 8 using the de-trended data. The fluences of the GLE events from NM data were not examined in the Oh et al. (2010) analysis. Instead, they compared the fluences of the GOES events with the maximum increase in the NM data. Where their fluence was calculated as the time integrated intensity during the half day interval after the peak. For the P8 and P10 channel they found coefficients of 0.821 and 0.871, respectively. Equivalent calculations for our data set return a correlation coefficient of 0.742 and 0.813 for Ch 12 and 14.

GLE 69 is the most extreme GLE in our sample, with a NM increase at least an order of magnitude larger than the rest of the GLEs. Its extreme nature and properties have already been well documented and studied. For example, GLE 69 was associated with a gamma ray flare and observed to have an extremely fast rise and strong anisotropy during the impulsive peak of the event (Grechnev et al., 2008; Klein et al., 2014). GLE 69 appears as an outlier in many of our results and does not seem to fit as well with the rest of the sample. GLE 69s large NM increase and peak flux separates it from the main GLE cluster in Figure 8. The extreme nature of this "super GLE" has been suggested to in part be due to its location on the Western hemisphere, the favorable location of Earth's footpoint, as well as the proximity of the source location to the HCS at the time of the event (Waterfall et al., 2022).

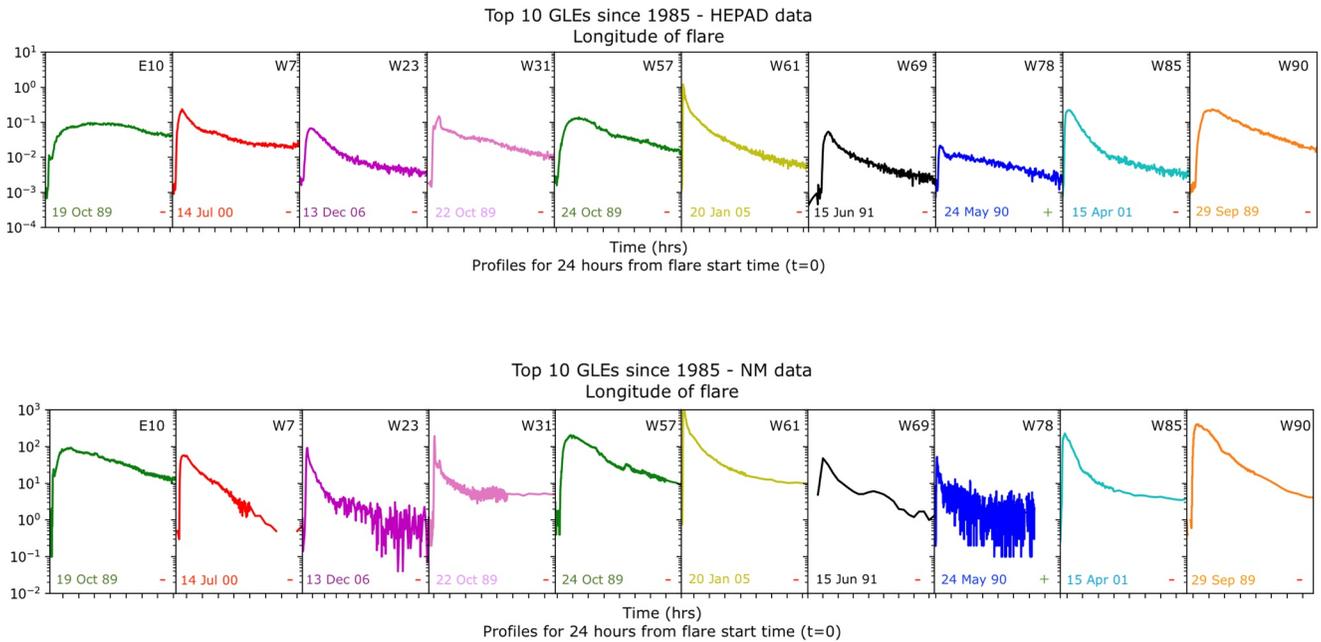

**Figure 9.** Top 10 ground level enhancements (GLEs) since 1985, ordered by longitude of associated flare. Top panel: solar energetic particle flux profiles, bottom panel: GLE event profiles from neutron monitors with largest increase. Event profiles are shown for 24 hr started from the associated flare start time.

The onset times of the GOES events compared to the NM profiles was also examined. The onset times in NM data were earlier or within 10 min of the GOES onset time in every case, consistent with higher energy particles causing GLE events. The time between flare start and onset in Ch14 is well correlated (0.75) with the time between flare start and onset in NM data (where onset time is determinable).

In addition to individual properties of these events, the profiles in general can be compared. Figure 9 shows the GOES (top) and NM (bottom) profiles of the 10 GLEs with the largest fluxes since 1985. These profiles are ordered by longitude, with GLE 43 on the far left with an Eastern longitude and GLE 42 on the right with the Western most longitude. It is clear from this ordering that among the largest GLEs, there is not a clear pattern of profile type versus longitude. One of the "lowest" fluxes in this set is for the 15 June 1991 GLE which was favorably located at W69. Figure 9 shows that the NM and high energy GOES profiles have a high degree of similarity. The shape of the rise and decay profile agree for many of the events. Again, GLE 69 (20 January 2005) is notable for its strikingly sharp impulsive component and profile shape.

## 7. High Energy Events and the Heliospheric Current Sheet

When the polarity of the heliospheric magnetic field during the event is examined we find that 9 out of 10 of these strongest GLEs occur during an A− configuration (i.e., solar magnetic field pointing inward in the northern hemisphere). A similar pattern was seen in Waterfall et al. (2022) where only GLEs with percentage increases above 15% were examined. Of the 17 GLEs in their event set, 14 occurred during an A− configuration. The majority of these GLEs were found to be affected by the proximity of the HCS to the source location. 3D modeling of the transport of energetic solar protons by Dalla et al. (2020) also found that fluxes were lower during A+ periods. They suggested this was due to the polarity of the field affecting drift motion near the HCS, with an A+ configuration causes more direct and faster transport along the HCS toward the outer heliosphere. When looking at the polarity for other events in the sample not included in Figure 9, it was predominately A+ for the events of solar cycle 24. No solar cycle 24 GLEs are present in Figure 9 and have very small NM increases. The trend of lower peak fluxes for SEP events that occur during an A+ configuration is seen in Figure 10. The flare intensities and Ch12 peak fluxes are the same as in Figure 5. For a given flare magnitude, the smaller Ch12 events are generally A+ events.

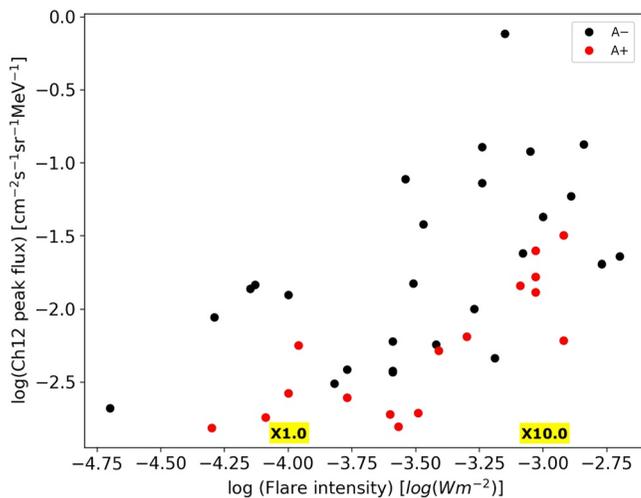

**Figure 10.** Peak SXR flare intensities and Ch12 fluxes from the 42 solar energetic particle events (as in Figure 5). The events that occur during an A+ configuration are highlighted in red. The correlation coefficients for the A+ and A− populations are 0.80 and 0.56, respectively (an increase from 0.55 in Figure 5).

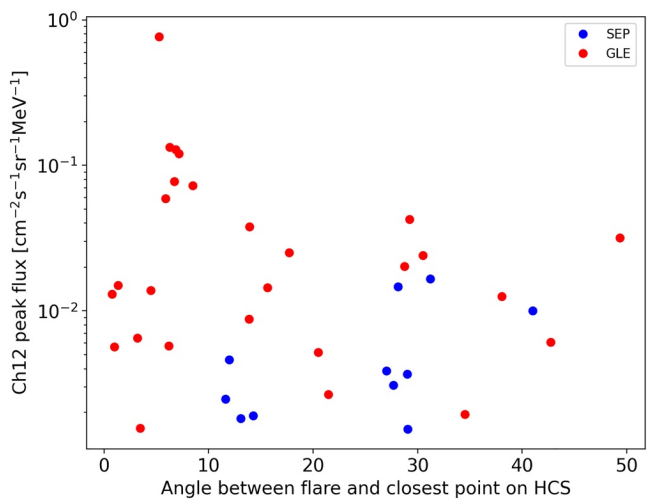

**Figure 11.** Plot of the peak solar energetic particle (SEP) flux measured in Ch12 versus the angular distance between the flare's location and closest point on the heliospheric current sheet. Ground level enhancement (GLE) events are shown in red, non-GLE SEP events in blue.

Locations of large flares associated and not-associated with GLEs and their distance to the HCS have previously been studied by Waterfall et al. (2022), for the period between 1976 and 2020. They found that 44% of GLE-associated source regions were located less than 10° from the HCS. Figure 11 considers how the peak flux varies with angular distance between the flare location and HCS for the events in the current study. This angle is calculated from source surface map data, obtained from the Wilcox Solar Observatory (WSO) at http://wso.stanford.edu. The maps contains information on the HCS at the time of the event and are built using potential field modeling with photospheric magnetogram data. There is a weak negative correlation (coefficient of −0.23 for Ch12) between the angle and peak flux for all events. Further exploring potential differences between the GLE and non-GLE populations, the events are color coded red (GLE) and blue (non-GLE). We find that all events with angles less than 10° are GLEs, making up 52% of the total GLEs. Additionally, all events with peak fluxes $>2 \times 10^{-2}$ cm$^{-2}$ s$^{-1}$ sr$^{-1}$ MeV$^{-1}$ in Ch12 are GLEs.

## 8. Discussion and Conclusions

We have conducted an analysis of 42 high energy SEP events observed during the period 1984–2017, covering three complete solar cycles. Several features of the associated flare, CME and GLE event are compared with high energy GOES data and several relationships are found. These relationships are compared to existing analysis of lower energy SEP events and NM studies. Our main conclusions are as follows:

- The majority of SEP events observable at energies >300 MeV are associated with GLE events (29/42).
- The correlations seen between the flare magnitude and peak flux in all three high energy channels are generally weaker than those seen for lower energy analysis. In contrast, the correlations for fluence are comparable. These coefficients are compared in Table 4. Correlations for peak flux are improved when the coefficients are calculated according to the IMF polarity at the time of the event.
- Correlations between the CME velocity and peak flux (and fluence) are weak in all channels, in contrast to moderate correlations seen in lower energy analysis. This conclusion is based on a subset of our events for which CME data were available.
- There are strong correlations and similarities seen between the properties of the SEP event and NM data for GLEs. The strongest correlation (>0.9) is seen between the peak flux and NM maximum increase for all high energy channels.
- The GOES and NM profiles for the largest events also share similar features, despite the GLE particles being affected by the Earth's magnetic field and the location of the NM
- The larger GLE intensity profiles are not strongly ordered by location of the associated flare, as is often the case for low energy SEPs. This may be due to effects such as HCS drift (Waterfall et al., 2022) and gradient and curvature drifts (Dalla et al., 2013, 2020) being more important at these energies.

We have calculated many features of our high energy data set, including; peak flux in three high energy channels, event fluence, onset time, duration, etc. These have all been compared to the associated flare longitude, peak SXR flux, SXR onset time and duration, CME speed, CME width, etc. We have reported here on the most significant of these relationships, and those which stray from results from lower energy analysis. The strongest correlation was found between the flare magnitude and the Ch12 fluence at 0.62 when examining the flare parameters. Analysis of lower energy GOES data between a similar time period by Papaioannou et al. (2016) obtained coefficients of 0.48 and 0.42 for >10 and >100 MeV. Conversely, our results from comparing the CME velocity with the peak fluxes do not produce comparable results to low energy analysis. Papaioannou et al. (2016) obtain correlation coefficients that decrease from 0.57 to 0.40 from >10 to >100 MeV. We see much weaker correlations across our channels (lowest at 0.14 for Ch13). In the CME analysis we were limited by the reduction of our already small event list due to the lack of CME data in the 1980s. Despite this, the results suggest that for high energies a faster CME does not lead to an increase in higher energy particles. CME data have previously been used to derive an empirical model to predict SEP spectra for 10–130 MeV (Bruno & Richardson, 2021). SEP spectra are an important feature that may help in understanding the correlations seen at higher energies in this study. However, an in-depth look at SEP spectra is currently beyond the scope of this work but is worthwhile for future studies.

We are limited in our analysis by the relatively few number of >300 MeV SEP events compared to their lower energy counterparts and the availability of spacecraft that have been operational long enough to observe multiple high energy events. Analysis has previously been performed on extreme SEP events observed by PAMELA (Bruno et al., 2018), however, only SEP events that occurred between 2006 and 2016 are available. This time period was relatively quiet in terms of solar activity and only two GLEs occurred. Thirty SEP events are used in that study, where six events have an associated source behind the limb (a feature of the excluded events in this analysis) and only 6 of the remaining 24 events were also analyzed in this high energy study. While there are limited high energy data sets from spacecraft, we can also make use of ground-level data from neutron monitors. Information on the peak increases and event durations during multiple GLEs detected by neutron monitors has been used in this analysis. The largest GLE events in the GOES data set all have well-defined and strong NM profiles. These NM profiles were compared to corresponding GOES profiles and several similarities were seen between profile shapes. Very strong correlations are seen between the maximum NM increase and peak Ch12 and Ch14 fluxes. Prior analysis by (Oh et al., 2010) also reported correlations between NM and GOES peaks. Strong correlations are seen between the fluences of the NM and GOES events across several orders of magnitude. Not every high energy SEP event observed by GOES (>300 MeV) is associated with a GLE (thought to be associated with protons of energy >500 MeV). For the events that are, the strong correlations suggests that: acceleration and propagation at >300 and >500 MeV are similar, and that the effects of the geomagnetic field preserve peak intensities and overall shape, at least at the NM that observed the highest fluxes (usually in polar regions).

One notable GLE in our data set is GLE 69 (20 January 2005). This GLE has been extensively studied in literature because of its very impulsive profile and peak flux (and NM increase) orders of magnitude larger than other SEP events. Despite fitting with many of the correlations and trends seen in this paper, GLE 69s large flux is frequently seen as an outlier in many of our plots. We refer to it as a "super-GLE" due to this, however note that its associated flare is large (X7.1) but still comparable to the other events. Its associated CME was very fast, with a reported speed of 3,256 km s$^{-1}$ (Papaioannou et al., 2016). Previous work by Lario et al. (2008) has explored the possible existence of a streaming limit for SEP events. However, the existence of GLE 69 and its extreme peak flux values poses a challenge to the streaming limit concept. There are many factors that influence a solar eruptive events effect at a given heliospheric location (e.g., source longitude, CME speed, peak X-ray flare intensity), and it can be argued that many of these factors perfectly aligned at the time of this event. While no GOES data exists for GLE 5 in 1956, its NM increase (larger than that of GLE 69) and the strong correlation seen here between peak flux and NM increase further supports the idea that the streaming limit does not exist and even larger peak fluxes are possible.

The events in our sample have a wide range of flares and CME speeds (M2.0–X20 and 792–3,256 km s$^{-1}$, respectively). Flares and CMEs of a similar size have been observed that do not lead to SEP events. One common explanation for SEP events not occurring is the Eastern location, and thus poor magnetic connection with Earth, of the particles source location. However, of the 13 non-GLE SEP events in Table 2, only 4 have Eastern longitudes. There have been other suggestions as to why some events are larger than others, for example, curvature and gradient drifts induced by the Parker spiral or proximity to the source region to the HCS (Dalla et al., 2013; Waterfall et al., 2022). However, more data is needed to explore this further. Time will provide more high energy SEP events, but we need the detectors to analyze them. Understanding why some large solar eruptive events cause SEP events (or indeed GLE events) and some do not is crucial to space weather research and improving forecasts. Statistical studies of SEP events are often utilized in forecasting models. For example, correlations between solar flare parameters and SEP peak fluxes are currently used in the SEP forecasting model SPARX (Marsh et al., 2015). Other parameters, such as CME speeds, are not used in this model as they are not available in real-time. However, future observations and inclusion of features like these will aid in refining these forecast models.

In general, there is a lack of comprehensive datasets that cover energies >300 MeV. Events at these energies are arguably the most important to space weather related hazards and understanding them more is vital. GOES-HEPAD remains the only detector that provides observations over multiple decades up to several hundred MeV. Similar high energy detectors, such as GME–IMP8 has many data gaps and was unusable in this study. As space exploration continues to increase, there is a need for more information about these high energy events. Such information and analysis of high energy data can be used in forecasting models for the aviation industry as well as to help protect astronauts and infrastructure in space.

## Data Availability Statement

This manuscript has used the SEPEM reference data set version 3 (Jiggens et al., 2022), which has been extended to high energies using re-calibrated HEPAD data (Raukunen et al., 2020). The data was originally obtained from GOES observations which were cleaned, background subtracted, and cross-calibrated (Rodriguez et al., 2017; Sandberg et al., 2014). The original data is available from: https://satdat.ngdc.noaa.gov/sem/goes/data/avg/. Neutron monitor data was obtained from the GLE database https://gle.oulu.fi, managed and hosted by the Oulu Cosmic Ray Station at the University of Oulu, Finland (Usoskin et al., 2020). Information on the heliospheric current sheet was obtained from the Wilcox Solar Observatory (WSO), currently operated by Stanford University with funding provided by the National Science Foundation. WSO data is accessible via http://wso.stanford.edu. Information on existing SEP events was obtained from the NOAA Space Environment Services centre list at: https://umbra.nascom.nasa.gov/SEP/. Figures were created using Matplotlib version 3.3.2 (Hunter, 2007), available from https://matplotlib.org.

**Acknowledgments**

C.O.G. Waterfall and S. Dalla acknowledge support from NERC via the SWARM project, part of the SWIMMR programme (Grant NE/V002864/1). SD acknowledges support from the UK STFC through grants ST/R000425/1 and ST/V000934/1 and from the International Space Science Institute through funding of the International Team on "Solar Extreme Events: Setting up a paradigm." We acknowledge the use of data from Wilcox Solar Observatory data in this study.